\documentclass[a4paper]{jpconf}
\usepackage{graphicx}
\usepackage{amsfonts}
\begin{document}
\title{Weak Field Limit of the Nonminimally Coupled Weyl Connection Gravity}

\author{Cláudio Gomes}

\address{Centro de Física das Universidades do Minho e do Porto, Faculdade de Ciências da Universidade do Porto, Rua do Campo Alegre s/n, 4169-007 Porto, Portugal}
\address{Instituto Okeanos, Universidade dos Açores, Rua Prof. Doutor Frederico Machado 4, 9901-862 Horta, Faial, Açores, Portugal}

\ead{claudio.gomes@fc.up.pt}

\begin{abstract}
The true nature of gravity is a remarkable open problem in Gravitation. Theoretical and observational motivations open the avenue of alternative theories of gravity. One possibility resorts to nonminimal couplings and non-metricity properties of spacetime, and is dubbed as nonminimally coupled Weyl connection gravity. It has the advantage of leading to metric field equations of second order together with a constraint equation for the Weyl vector, and has well behaved space-form. We analyse this model by exploring its weak regime and its implications for astrophysics and cosmology.
\end{abstract}

\section{Introduction}

Dark matter and dark energy are two puzzling problems in Modern Cosmology. The missing mass on galaxies and clusters and the recent acceleration phase of the Universe can occur either from exotic fields and fluids or from modified gravity models. The former approach has been proved unsuccessful in findings such components, while the latter may have additional advantages in what concerns to deal with the cosmological problem, or hints towards an unification of the four fundamental interactions. Moreover, the best gravitational theory until now is Einstein's General Relativity (GR) which favourably explains a profusion of phenomena and meets the stringent solar system tests, namely the Mercury's perihelion precession, the Shappiro's time delay, the bending of light in gravitational fields, and the gravitational redshift, as well as several other data. Nonetheless, in addition to the aforementioned dark matter and dark energy conundrums at large scales, GR presents singularities, lacks a fully consistent quantum version, has the cosmological constant problem, among other conundrums.

In fact, there is an extensive literature on alternative gravity models, from which the simplest one is the well-known f(R) theories \cite{fr1,fr2} in which the Ricci scalar from the GR action functional is replaced by a generic function of it. Moreover, one may further generalise such scenario by including a non-minimal coupling between the matter Lagrangian and an additional function of the curvature scalar in what is known as the non-minimal matter-curvature coupling model \cite{nmc}. This model allows for reproducing effective background expansion histories in the brane-world scenario and in and loop quantum cosmology approach (resorting to both metric and Palatini formalisms, where the latter exhibits advantages over the former) \cite{olmo}, for the $\Lambda$CDM paradigm to be suitably embedded \cite{mariam}, and for a resemblance with effective gravity expanded around a Minskowskian background, as one generally thinks on the free functions as a power-law series sum, $f_i(R)=\sum_n a_n \left(\frac{R}{R_n}\right)^n$, with $i=1,2$ and $n\in \mathbb{Z}$.

Nonetheless, these and other approaches are made with the metric satisfying the metric condition, $\nabla_{\mu}g^{\mu\nu}=0$, and in the absence of torsion. One can prove that there is a geometric trinity in the sense that one can build quantities for torsion and non-metricity such that their theories are equivalent to General Relativity up to boundary terms \cite{jimenez}. However, when one attempts to generalise both scenarios to a generic function of those quantities, then they are not equivalent to each other nor with $f(R)$ theories. This degeneracy breaking is often seen as an opportunity to understand the true nature of these gravity properties.

Thus, one may address a further non-minimal coupling version of the previous cases, namely with non-metricity \cite{nmcfQ}. The non-metricity can have different formulations, and can also coexist with the metric field with both defining the spacetime. One of such examples is the Weyl gravity \cite{weyl}, which was an attempt to unify electromagnetism to General Relativity. Dirac has reformulated this scenario of describing spacetime by considering a new scalar field in addition to the metric field \cite{dirac}. Despite early problems from its original formulation, alternative forms for the Weyl gravity have been formulated and revisited as explanations for the dark energy and dark matter issues, and for the inflation paradigm \cite{alvarez}.

In fact, one may even resort to solely the Weyl connection in the non-minimal coupling framework leading to metric field equations of second order together with a constraint equation for the Weyl vector \cite{nmcweyl}. This model admits a viable cosmological description \cite{baptista2} and when it is dynamical and identified as a electromagnetic potential-like vector Ostragradsky instabilities can be avoided provided the Weyl vector obeys the form $A_{\mu}=(0,A(r),0,0)$ or the extrinsic curvature scalar of the hypersurface of the space-time foliation vanishes \cite{baptista2}. Furthermore, black hole solutions were found and characterised in Ref. \cite{blackholes}.

The goal of the present work is, therefore, to analyse the weak field limit of the nonminimally coupled Weyl connection model, and its implications.

\section{The Basics on the Weyl Connection Model}

The Weyl Connection model is characterised by the generalisation of the covariant derivative with respect to a more general connection related to the Weyl vector, $A_{\lambda}$, which upon acting on the metric field tensor, $g_{\mu\nu}$, extracts the Weyl vector in the following way:
\begin{equation}
	D_\lambda g_{\mu\nu}=A_\lambda g_{\mu\nu},
\end{equation}
from which the generalised covariant derivative reads: 
\begin{equation}
	D_\lambda g_{\mu\nu}=\nabla_\lambda g_{\mu\nu}-\bar{\bar{\Gamma}}^\rho_{\mu\lambda}g_{\rho\nu}-\bar{\bar{\Gamma}}^\rho_{\nu\lambda}g_{\rho\mu},
\end{equation}
which is a sum of the Levi-Civita covariant derivative $\nabla_\lambda$ with the additional terms on the connection that arise from the non-metricity $\bar{\bar{\Gamma}}^\rho_{\mu\nu}=-\frac{1}{2}\delta^\rho_{\mu}A_\nu-\frac{1}{2}\delta^\rho_\nu A_\mu+\frac{1}{2}g_{\mu\nu}A^{\rho}$.

Therefore, the curvature (actually, the change of shape of an element along geodesics) is now characterised by a generalised version of the Riemann tensor, $\bar{\Gamma}^\rho_{\mu\nu}=\Gamma^\rho_{\mu\nu}+\bar{\bar{\Gamma}}^\rho_{\mu\nu}$, such that
\begin{equation}
	\bar{R}^\rho_{\mu\sigma\nu}=\partial_\sigma \bar{\Gamma}^\rho_{\nu\mu}-\partial_\nu\bar{\Gamma}^\rho_{\sigma\mu}+\bar{\Gamma}^\rho_{\sigma\lambda}
	 \bar{\Gamma}^\lambda_{\nu\mu}-\bar{\Gamma}^\rho_{\nu\lambda}\bar{\Gamma}^\lambda_{\sigma\mu}.
\end{equation}

By contracting the first and third indices of this curvature tensor, one gets a generalised Ricci tensor that measures the change of volume along geodesics:
\begin{equation}
	\bar{R}_{\mu\nu}=R_{\mu\nu}+\frac{1}{2}A_\mu A_\nu +\frac{1}{2}g_{\mu\nu} \left(\nabla_\lambda-A_\lambda\right)A^\lambda +F_{\mu\nu}+\frac{1}{2}\left(\nabla_\mu A_\nu+\nabla_\nu A_\mu\right)=R_{\mu\nu}+\bar{\bar{R}}_{\mu\nu},
	\label{Riccitensor}
\end{equation}
where $R_{\mu\nu}$ is the standard Ricci curvature tensor and the strength tensor of the Weyl field is given by $F_{\mu\nu}=\partial_\mu A_\nu-\partial_\nu A_\mu=\nabla_\mu A_\nu -\nabla_\nu A_\mu$. One can compute its trace, leading to the scalar curvature:
\begin{equation}
	\bar{R}=R+3\nabla_\lambda A^\lambda-\frac{3}{2}A_\lambda A^\lambda=R+\bar{\bar{R}},
\end{equation}
where $R$ is the standard Ricci scalar.

Therefore, in the Weyl connection model the spacetime is described by both the metric and the Weyl vector. Moreover, the usual maximally symmetric Minkowski, de Sitter and anti-de Sitter spacetimes only match the constant Riemann curvature solutions if the Weyl vector field vanishes everywhere. In fact, this model can be formulated in what is known as Weyl gravity which differs from this by introducing a squared Weyl tensor in the action functional instead of the curvature scalar.

\section{The Nonminimally Coupled Weyl Connection Gravity Model}

We start by the action functional \cite{nmcweyl}:
\begin{equation}
S=\int \left[f_1(\bar{R})+f_2(\bar{R})\mathcal{L}\right]\sqrt{-g}d^4 x ~,
\end{equation}
where both $f_1, ~ f_2$ are functions of the scalar curvature, $\mathcal{L}$ is the Lagrangian density of matter fields, and $g$ is the determinant of the metric field.

The action can be varied with respect to the vector field leading to constraint-like equations \cite{nmcweyl}:
 \begin{equation}
 	\nabla_\lambda \bar{\Theta}(\bar{R})=-A_\lambda \bar{\Theta}(\bar{R}), 
 	\label{eqconstraint}
 \end{equation}
 where $\bar{\Theta}(\bar{R})=F_1(\bar{R})+F_2(\bar{R})\mathcal{L}$ and $F_i(\bar{R})=\frac{d f_i(\bar{R})}{d \bar{R}}$ , for both $i\in \{1,2\}$. 

Similarly, variation of the action with respect to the metric field, and resorting to the previous constraint equation, leads to \cite{nmcweyl}
 \begin{equation}
 	\left( R_{\mu\nu}+\bar{\bar{R}}_{(\mu\nu)} \right) \bar{\Theta}(\bar{R})-\frac{1}{2}g_{\mu\nu}f_1(\bar{R})=\frac{f_2(\bar{R})}{2}T_{\mu\nu},
 	\label{fieldeqs}
 \end{equation}
where we defined $\bar{\bar{R}}_{(\mu\nu)}=\frac{1}{2}A_\mu A_\nu+\frac{1}{2}g_{\mu\nu} \left(\nabla_\lambda-A_\lambda\right)A^{\lambda}+\nabla_{(\mu}A_{\nu)}$ and $T_{\mu\nu}$ is the energy-momentum tensor built of matter fields defined as in General Relativity as $T_{\mu\nu}=-\frac{2}{\sqrt{-g}}\frac{\delta (\sqrt{-g}\mathcal{L})}{\delta g^{\mu\nu}}$. 

One trace the metric field equations:
\begin{equation} 
\bar{\Theta}(\bar{R}) \bar{R}-2f_1(\bar{R})=\frac{f_2(\bar{R})}{2}T,
    \label{trace}
\end{equation}
where $T=g^{\mu\nu}T_{\mu\nu}$ is the trace of the energy-momentum tensor. 

Working around the previous equations, trace-free equations can be obtained:
\begin{equation}
   \bar{\Theta}(\bar{R}) \left[R_{\mu\nu}-\frac{1}{4}g_{\mu\nu}R \right]+\bar{\Theta}(\bar{R}) \left[\bar{\bar{R}}_{(\mu\nu)}-\frac{1}{4}g_{\mu\nu}\bar{\bar{R}} \right]=\frac{f_2(\bar{R})}{2}\left[T_{\mu\nu}-\frac{1}{4}g_{\mu\nu} T\right]. 
\label{trace-free_eqs}
\end{equation}

The covariant conservation laws from General Relativity are no longer valid as the divergence of the modified Einstein's field equations results in a covariant non-conservation law of the energy-momentum tensor \cite{nmcweyl}:
\begin{equation}
	\nabla_\mu T^{\mu\nu}=\frac{2\kappa}{f_2(\bar{R})}\left[ \frac{F_2(\bar{R})}{2}\left( g^{\mu\nu}\mathcal{L}-T^{\mu\nu} \right)\nabla_\mu R+\nabla_\mu(\bar{\Theta}(\bar{R}) B^{\mu\nu})-\frac{1}{2}\left(F_1(\bar{R})g^{\mu\nu}+F_2(\bar{R})T^{\mu\nu}\right)\nabla_\mu \bar{\bar{R}} \right],
	\label{nonconserveq}
\end{equation}
where $B^{\mu\nu}=\frac{3}{2}A^\mu A^\nu+\frac{3}{2}g^{\mu\nu}(\nabla_\lambda-A_\lambda)A^{\lambda}$. Thus, a non-trivial exchange of energy and momentum between matter and curvature occurs due to both non-minimal coupling and non-metricity.

\subsection{The Space-form Behaviour}
One can define a space-form as a Riemannian manifold for which its sectional curvature is constant. In other words, a space-form has the Riemann tensor components obeying the following:
\begin{equation}
R_{abcd}=K\left(g_{ac}g_{db}-g_{ad}g_{cb}\right)~,
\end{equation}
where $K$ is some real constant.

This implies that the Ricci tensor reads $R_{bd}=3Kg_{bd}$ and, hence the scalar curvature $R=12K$. The constant $K$ is associated with the cosmological constant in General Relativity and signals a well defined vacuum state.

It is straightforward to generalise this concept to our case by replacing the Levi-Civita built curvature quantities by the Weyl connection analogous quantities. Furthermore, one may relax the time-independence of the vector field.

Two possible ansätze for the Weyl vector field that do not tarnish the Universe's homogeneity and isotropy are the following \cite{vector1,vector2}:
\begin{eqnarray}
&&A_{\mu}^{(1)}=A_0(t)\delta^0_{\mu}\\
&&A_{\mu}^{(2)}=A_1(t)\delta^a_i L_a ~,
\end{eqnarray}
where the generators of the internal $SO(3)$ group are denoted by $L_a$, with indices $a=1,2,3$. The first scenario leads to $A_0(t)=\frac{2A_0(t_0)}{2+A_0(t_0)(t-t_0)}$, with $t_0$ some initial instant. The second one leads to the trivial result $A_1(t)=0$ \cite{nmcweyl}.

\section{The Newtonian Regime}

We now proceed to compute the Newtonian regime of the nonminimally coupled Weyl connection gravity model. We draw the attention to the fact that Newtonian and Post-Newtonian approximations lead to different results as the latter considers higher corrections in the expansion series of inverse powers of the speed of light \cite{capozzielloweak}. The first approach is the one used to compute the Jeans' criterion for gravitational instability \cite{binney,capozziellojeans,jeans1,jeans2}. The second leads to the study of the Yukawa potential in the Solar System analysis. This is more evident in numerical simulations \cite{numerics1,numerics2,numerics3}.

We shall expand the following quantities up to order $c^{-2}$:
\begin{eqnarray}
&&g_{\mu\nu}\approx\eta_{\mu\nu}+h_{\mu\nu} ~,\\
&&\bar{R}\approx R^{(2)}\equiv \delta R ~,\\
&&f_i^n(\bar{R})\approx f_i^{n}(0)+f_i^{n+1}(0)\delta\bar{R}~,\label{eqn:vectorchoice}
\end{eqnarray}
where $|h_{\mu\nu}| \ll 1$.

Thus, up to $\mathcal{O}(c^{-2})$ the time-time component of the metric field equations and the trace equation can be expanded around the Minskowski spacetime where at lowest order matter fields do not contribute:
\begin{eqnarray}
&&\left(\delta R_{00}+\delta \bar{\bar{R}}_{(00)}\right)F_1(0)-\frac{1}{2}\eta_{00}F_1(0)\delta\bar{R}=\frac{f_2(0)}{2}\delta T_{00} ~,\\
&& -F_1(0)\delta\bar{R}=\frac{f_2(0)}{2}\delta T~, 
\end{eqnarray}
because the lowest order (zeroth order) of the expansion implies $f_1(0)=0$. In fact, these first order perturbations act on the energy-momentum tensor, such that for a perfect fluid, it should behave as $\delta T_{\mu\nu}=\rho\delta_{\mu}^0\delta_{\nu}^0$ and $\delta T=-\rho$. Furthermore, the metric behaves as $g_{\mu\nu}=diag(-1-2\Phi,1-2\Psi,1-2\Psi,1-2\Psi)$, where $\Phi,~\Psi$ are the usual Bardeen scalar potentials that obey $|\Phi|,|\Psi|\ll 1$. This implies that the perturbed Ricci tensor and Ricci scalar read $\delta R_{00}=\nabla^2\Phi$ and $\delta R=2\nabla^2(2\Psi-\Phi)$.

Thus:
\begin{eqnarray}
&&\left(\nabla^2\Phi+\delta \bar{\bar{R}}_{(00)}\right)+\frac{1}{2}\delta\bar{R}=\frac{\gamma}{2}\rho ~,\label{eqn:poisson1}\\
&& \delta\bar{R}=\frac{\gamma}{2}\rho~,\label{eqn:poisson2}
\end{eqnarray}
where $\alpha\equiv F_1'(0)/F_1(0)$, $\beta\equiv F_2(0)/F_1(0)$ and $\gamma\equiv f_2(0)/F_1(0)$.

We now proceed to choose suitable ans\"{a}tze for the Weyl vector field. We can start by inspecting those in the form of the ones compatible with the space-form behaviour, noting however, that in our case they should depend on the distance and not very much on time.

\subsection{First Ansatz: $A_{\mu}= (A_0(x,y,z),0,0,0)$}
The first ansatz we shall use is compatible with spherically symmetric solutions and is of the form \cite{nmcweyl}: 
\begin{equation}
A_{\mu}\approx (A_0(x,y,z),0,0,0)~,
\end{equation}
where we shall require $~|A_0(x,y,z)|\ll 1$.

Thus, the constraint equation can also be expanded around a Minkowskian spacetime:
\begin{equation}
\nabla_{\lambda}\left[F_1'(0)\delta\bar{R}+F_2(0)\delta\mathcal{L}\right]=-A_0\delta^0_{\lambda}F_1(0)~.
\end{equation}

Hence, this equation can be rewritten resorting to the parameters $\alpha,\beta,\gamma$ previously defined and together with Eqs. (\ref{eqn:poisson1}) and (\ref{eqn:poisson2}), we get a system of equations:
\begin{eqnarray}
&&\left(\nabla^2\Phi+\delta \bar{\bar{R}}_{(00)}\right)+\frac{1}{2}\delta\bar{R}=\frac{\gamma}{2}\rho ~,\\
&& \delta\bar{R}=\frac{\gamma}{2}\rho~,\\
&& -\nabla_{0}\left[\alpha\delta\bar{R}-\beta\rho\right]=-A_0
\end{eqnarray}

Plugging the equation in the middle into the other two, and explicitly expanding the one in the middle in terms of full expression for the scalar curvature:
\begin{eqnarray}
&&\left(\nabla^2\Phi+\delta \bar{\bar{R}}_{(00)}\right)=\frac{\gamma}{4}\rho ~,\\
&& \delta\bar{\bar{R}}+2\nabla^2(2\Psi-\Phi)=\frac{\gamma}{2}\rho~,\\
&& \left(\frac{\alpha\gamma}{2}-\beta\right)\dot{\rho}=A_0~, 
\end{eqnarray}
where the dot derivative stands for a time derivative. Since in the weak field limit, time derivatives are small and can be neglected at first order, and since we expect that $\alpha,\beta,\gamma \sim \mathcal{O}(1)$, then $A_0$ must vanish at first order, hence the Weyl vector field in the form of Eq. (\ref{eqn:vectorchoice}) may only impact in a post-Newtonian approximation. In the non-appearance of the Weyl vector, the metric field equations reduce to the well known case of the non-minimal matter-curvature coupling model.

\subsection{Second Ansatz: $A_{\mu}= (A_0,A_1,A_1,A_1)$}
The second choice for the Weyl vector field is the most general expression compatible with a spherically symmetric physical system (when rewritten in spherical coordinates), namely:
\begin{equation}
A_{\mu} = (A_0(x,y,z),A_1(x,y,z),A_1(x,y,z),A_1(x,y,z))~,
\end{equation}
with $|A_0|,|A_1| \ll 1$. In this case, we get the following system of equations:
\begin{eqnarray}
&&\left(\nabla^2\Phi+\delta \bar{\bar{R}}_{(00)}\right)+\frac{1}{2}\delta\bar{R}=\frac{\gamma}{2}\rho ~,\\
&& \delta\bar{R}=\frac{\gamma}{2}\rho~,\\
&& -\nabla_{0}\left[\alpha\delta\bar{R}-\beta\rho\right]=-A_0~,\\
&& \nabla_{j}\left[\alpha\delta\bar{R}-\beta\rho\right]=-A_1~.
\end{eqnarray}

By the same reasoning as the former case, for this ansatz, we conclude that the only solution is a null time component for the Weyl vector field. Thus, and neglecting terms of $\mathcal{O}(A^2)$, this system can be rewritten as:
\begin{eqnarray}
&&\nabla^2\Phi-\frac{1}{2}\nabla_{\lambda}A^{\lambda}=\frac{\gamma}{4}\rho ~,\\
&& 3\nabla_{\lambda} A^{\lambda}+2\nabla^2(2\Psi-\Phi)=\frac{\gamma}{2}\rho~,\\
&& A_0 = 0,\\
&& \left(\frac{\alpha\gamma}{2}-\beta\right)\nabla_{j}\rho=-A_1~.
\end{eqnarray}

Hence, plugging the second equation of this system into the first one, and applying this solution back on the second equation, we get:
\begin{eqnarray}
&&\nabla^2(\Phi-\Psi)=\nabla_{\lambda}A^{\lambda} ~,\\
&& \nabla^2(\Psi+\Phi)=\frac{\gamma}{2}\rho~,\\
&& A_0 = 0,\\
&& \left(\frac{\alpha\gamma}{2}-\beta\right)\nabla_{j}\rho=-A_1~.
\end{eqnarray}

Moreover, the fourth equation can be put into the first one, such that:
\begin{eqnarray}
&&\left(\frac{\alpha\gamma}{2}-\beta\right)\nabla^2(\Psi-\Phi)=\nabla^2\rho ~,\\
&& \nabla^2(\Psi+\Phi)=\frac{\gamma}{2}\rho~,\\
&& A_0 = 0,\\
&& \left(\frac{\alpha\gamma}{2}-\beta\right)\nabla_{j}\rho=-A_1~.
\end{eqnarray}

Finally, the fourth equation can be put into the first one, such that:
\begin{eqnarray}
&&\nabla^2\Psi=\left(\frac{\gamma}{2}+\frac{\nabla^2}{Z}\right)\rho ~,\\
&&\nabla^2\Phi=\frac{\nabla^2}{Z}\rho~,\\
&& A_0 = 0,\\
&& Z\nabla_{j}\rho=-A_1 \iff Z \nabla_{\lambda}\rho=-A_{\lambda}~,
\end{eqnarray}
with $Z= \left(\frac{\alpha\gamma}{2}-\beta\right)$.

Hence, we get a system that can be numerically solved provided an input on the energy density, for instance, is given.

\section{Conclusions}

We have examined a modified gravity model in which the non-minimal matter-curvature coupling model included the Weyl connection. This leads to second order metric field equations together with the Weyl vector satisfying a constraint equation.

This model as a well behaved space-form behaviour, and has been shown to be stable under cosmological solutions and on the Ostrogradsky analysis (where the Weyl vector field was associated with the electromagnetic 4-vector potential, leading to some constraints on the form of the vector field in this case).

Thus, the weak field regime of this gravity model was studied in this work. For the first ansatz with only a nonvanishing time component that depends on the distance coordinates, we found the only solution is a total vanishing Weyl vector field. Hence, in this case, the weak field limit of the theory corresponds to the one already studied in the standard non-minimal matter-curvature coupling gravity model. However, for a second ansatz with two distinct functions (one for the time component and another which was the same for all the spatial components dubbed as $A_1$), we got a vanishing time component and a set of equations giving explicit solution for the Bardeen's scalar gauge invariants and for $A_1$ provided we know the behaviour of the energy density.

Therefore, future work must peruse the post-Minskowskian limit of the theory, where the Weyl vector field may impact the solutions that will have both a Yukawa coupling and an extra-force contribution.

Moreover, a recent work on black hole solutions \cite{blackholes} showed some new features on the classical Schwarzschild and Reissner-Nordstrom-like exact solutions, and the need to revisit the first law of thermodynamics so to match both quantum tunnelling and area thermodynamical quantities. Therefore, further studies are necessary, namely analysing its consequences for the thermodynamics the Universe, since the metric non-minimal coupling between matter and curvature gravity model has some bearings in this context \cite{boltzmann}.

\ack
This work is funded by FCT – Fundação para a Ciência e a Tecnologia, through Project No. UIDB/04650/2020.

\end{document}